\documentclass[conf]{new-aiaa}
\usepackage[utf8]{inputenc}

\usepackage{graphicx}
\usepackage{amsmath}
\usepackage[version=4]{mhchem}
\usepackage{siunitx}
\usepackage{longtable,tabularx}
\usepackage{bm}

\newtheorem{rem}{Remark}

\newcommand{\intv}[2]{[#1{:}#2]}

\usepackage{algorithm}
\usepackage{algpseudocode}

\newcommand{\derv}[1]{\overset{{\scriptscriptstyle\circ}}{#1}}

\usepackage{graphicx}
\usepackage{bm}
\usepackage{xcolor}

\definecolor{Lyellow}{RGB}{255,255,220}

\usepackage{fancybox}

\newcommand{\greentick}{\textcolor{green!70!black}{\checkmark}}
\newcommand{\yellowcircle}{\textcolor{yellow!80!black}{$\bm{\circ}$}}

\newcommand{\purpletriangle}{\textcolor{purple!100!black}{$\triangle$}}
\newcommand{\redcross}{\textcolor{red}{$\times$}}

\newcommand{\shft}{c}

\definecolor{beige}{RGB}{245,245,230}

\usepackage[many]{tcolorbox}
\newtcolorbox{eqbox}{
  colback=beige!30,     
  colframe=black!30,    
  boxrule=0.7pt,        
  arc=5pt,              
  boxsep=0pt,           
  left=0pt, right=5pt,  
  top=6pt, bottom=5pt
}
\newtcolorbox{eqboxb}{
  colback=beige!30,     
  colframe=black!30,    
  boxrule=0.7pt,        
  arc=5pt,              
  boxsep=0pt,           
  left=0pt, right=5pt,  
  top=-6pt, bottom=5pt
}

\title{\Large Sequential Convex Programming for 6-DoF Powered Descent Guidance \\ with Continuous-Time Compound State-Triggered Constraints}

\author{Samet Uzun\footnote{Ph.D. Student, William E. Boeing Department of Aeronautics \& Astronautics; samet@uw.edu, AIAA Member.} and Beh{\c{c}}et A{\c{c}}{\i}kme{\c{s}}e\footnote{Professor, William E. Boeing Department of Aeronautics \& Astronautics; behcet@uw.edu, AIAA Fellow.}}
\affil{University of Washington, Seattle, WA 98195, USA}
\author{John M. Carson III\footnote{Technical Integration Manager – Precision Landing, NASA STMD; john.m.carson@nasa.gov, AIAA Fellow.}}
\affil{NASA Johnson Space Center, Houston, TX 77058, USA}

\begin{document}

\vspace*{-5.5em}

{\noindent
\begin{center}
    This is the accepted manuscript of a paper presented at AIAA SciTech Forum 2025. \\
    The published version is available at \href{https://doi.org/10.2514/6.2025-1895}{DOI:10.2514/6.2025-1895}.
\end{center}
}

\maketitle








%

\begin{abstract}
This paper presents a sequential convex programming (SCP) framework for ensuring the continuous-time satisfaction of compound state-triggered constraints, a subset of logical specifications, in the powered descent guidance (PDG) problem. The proposed framework combines the generalized mean-based smooth robustness measure (D-GMSR), a parameterization technique tailored for expressing discrete-time temporal and logical specifications through smooth functions, with the continuous-time successive convexification (CT-SCvx) method, a real-time solution for constrained trajectory optimization that guarantees continuous-time constraint satisfaction and convergence.
The smoothness of the temporal and logical specifications parameterized via D-GMSR enables solving the resulting optimization problem with robust and efficient SCP algorithms while preserving theoretical guarantees. In addition to their smoothness, the parameterized specifications are sound and complete, meaning the specification holds if and only if the constraint defined by the parameterized function is satisfied. The CT-SCvx framework is then applied to solve the parameterized problem, incorporating: (1) reformulation for continuous-time path constraint satisfaction, (2) time-dilation to transform the free-final-time PDG problem into a fixed-final-time problem, (3) multiple shooting for exact discretization, (4) exact penalty functions for penalizing nonconvex constraints, and (5) the prox-linear method, a convergence-guaranteed SCP algorithm, to solve the resulting finite-dimensional nonconvex PDG problem. The effectiveness of the framework is demonstrated through a numerical simulation. The implementation is available at \url{https://github.com/UW-ACL/CT-cSTC}
\end{abstract}
\vspace{-0.5cm}
\section{Introduction}
Autonomous guidance algorithms are an enabling technology for future manned and unmanned planetary missions.
Powered-descent guidance (PDG) involves generating reference trajectories and feedforward control commands to facilitate the soft landing of a rocket-powered vehicle on a planetary surface, playing an essential role in enabling human exploration missions to destinations within the solar system \cite{carson2019splice}.
Polynomial guidance methods developed for PDG \cite{sostaric2005powered, singh2007guidance, mendeck2011entry} have played a crucial role in the Apollo program for the landing of the lunar module on the Moon \cite{klumpp1974apollo} as well as in the Mars Science Laboratory (MSL) and Mars 2020 missions for the landing of the Curiosity and the Perseverance rovers on Mars \cite{san2013development, way2021reconstructed}.
While polynomial guidance methods have achieved remarkable success for PDG in space missions, the requirements for precise landing in challenging environments and the imposition of complex mission constraints have driven the development of more advanced PDG algorithms.

Advancements in numerical optimization algorithms have enabled the formulation and solution of the PDG problem as an optimal control problem.
The effectiveness of optimization algorithms on the PDG problem has been demonstrated in the landings of vertical-takeoff/vertical-landing reusable space rockets \cite{blackmore2016autonomous}.
The development of interior point methods \cite{Nesterov1994ipm, forsgren2002interior} has enabled the solving of convex optimization problems \cite{boyd2004convex} to global optimality, ensuring convergence within polynomial time.
Lossless convexification (LCvx) is a transformative method that applies convex optimization algorithms to solve the PDG problem \cite{acikmese2007convex}. Through the introduction of a slack variable, the nonconvex thrust magnitude lower bound constraint in the three-degree-of-freedom (3-DoF) PDG problem is transformed into a convex one. By leveraging Pontryagin's maximum principle, the convex problem's optimal solution is also optimal for the original nonconvex problem.
LCvx is subsequently extended to address a variety of constraints, including general nonconvex input sets \cite{accikmecse2011lossless}, minimum-error landing \cite{blackmore2010minimum}, thrust pointing constraints \cite{accikmecse2013lossless}, affine and quadratic state constraints \cite{harris2014lossless, harris2013lossless}, nonlinear dynamics \cite{blackmore2012lossless}, and binary constraints \cite{malyuta2020lossless}.
%
Customized second-order cone programming solvers \cite{dueri2014automated, dueri2017customized} are developed to assess the real-time performance of the LCvx, while the Guidance for Fuel Optimal Large Divert (G-FOLD) algorithm \cite{acikmese2013flight, scharf2014adapt, scharf2017implementation} was specifically developed to test the LCvx on Masten Space Systems' vertical-takeoff/vertical-landing suborbital rocket, Xombie \cite{xombie500, xombie750}.
Although considering only translation dynamics and constraints, related to either translation or the thrust vector via 3-DoF model, achieved sufficient success to solve the PDG problem, 6-DoF modeling of vehicle dynamics and increasing complexity of mission constraints necessitate the solution of more general nonconvex problems.

Sequential convex programming (SCP) has emerged as an effective method for transitioning from fully convex 3-DoF PDG problems to more complex, nonconvex 6-DoF PDG problems, which often include nonconvex constraints \cite{malyuta2022convex, malyuta2021advances}.
SCP is an iterative method that handles nonconvex problems by successively solving a sequence of locally convex approximations, achieved through linearizing the nonconvexities \cite{mao2016successive, reynolds2020real, bonalli2019gusto, cartis2011evaluation, drusvyatskiy2019efficiency}.
Therefore, SCP algorithms enable the handling of more general formulations of vehicle dynamics and mission constraints, which often introduce nonconvexities, such as aerodynamic forces on the vehicle \cite{szmuk2016successive}, free-final-time formulation \cite{szmuk2018successive}, multi-phase landing \cite{kamath2023seco}, state-triggered constraints \cite{szmuk2020successive, szmuk2019successive, reynolds2020dual}, and integer constraints \cite{malyuta2020fast, malyuta2023fast}.
The SCP algorithm has been further employed to solve quadrotor flight \cite{szmuk2017convexification, szmuk2018real, szmuk2019real} and hypersonic entry problems \cite{mceowen2022hypersonic, mceowen2023high, mceowen2024auto}, and its real-time capability has been demonstrated in quadrotor flight \cite{szmuk2017convexification, szmuk2018real, szmuk2019real} as well as PDG problems \cite{malyuta2019discretization, reynolds2020real}.
An SCP-based solution method utilizing the dual quaternion formulation for the PDG problem \cite{reynolds2020dual} has been selected as the candidate PDG algorithm for NASA's Safe and Precise Landing - Integrated Capabilities Evolution (SPLICE) project. This algorithm has been flight-tested in an open-loop configuration on the Blue Origin New Shepard suborbital rocket \cite{strohl2022implementation, doll2024performance}.
The traditionally utilized IPM algorithm to solve each convex subproblem in these algorithms is replaced with the customized first-order proportional-integral projected gradient (PIPG) algorithm \cite{yu2022proportional, yu2022extrapolated} for addressing the 3-DoF PDG problem via LCvx, leading to a significant improvement in the solution speed of the problem \cite{Elango2022}.
The PIPG algorithm is then employed to solve the 6-DoF PDG problem formulated via dual quaternion, resulting in a significant enhancement in real-time performance \cite{Kamath2023}.
In all these algorithms, path constraints are imposed only at the node points of the discretized continuous-time problem where inter-sample constraint satisfaction is not guaranteed. The continuous-time successive convexification framework (CT-SCvx), proposed in \cite{ctcs2024}, combines (i) a reformulation technique to ensure the satisfaction of the path constraints in continuous-time; 
(ii) time-dilation to transform the free-final-time optimal control problem into a fixed-final-time problem \cite{kamath2023seco};
(iii) exact penalty functions for penalizing nonconvex constraints \cite{nocedal2006numerical};
(iv) the multiple-shooting approach \cite{bock1984multiple, quirynen2015multiple} for an exact discretization of the continuous-time dynamics and (v) a convergence-guaranteed SCP algorithm, prox-linear method \cite{drusvyatskiy2018error, drusvyatskiy2019efficiency}, to solve the resulting finite-dimensional optimization problem. 
The implementations of the continuous-time successive convexification framework are demonstrated for passively-safe \cite{safe6dofpdg} and GPU-accelerated \cite{chari2024fast} trajectory optimization for 6-DoF PDG, nonlinear model predictive control (NMPC) for obstacle avoidance \cite{nmpc2024}, and trajectory optimization for 6-DoF aircraft approach and landing \cite{aircraft2024}.

While formulations exist to impose state-triggered constraints \cite{szmuk2020successive, szmuk2019successive, reynolds2020dual} and integer constraints \cite{malyuta2020fast, malyuta2023fast}, a general formulation has not yet been addressed for imposing temporal and logical specifications by ensuring their continuous-time satisfaction within the SCP framework.
Signal temporal logic (STL) specifications \cite{maler2004monitoring, donze2010robust} provide a general formulation for imposing temporal and logical specifications on optimization problems \cite{belta2019formal}; however, this framework involves $\min$ and $\max$ functions, leading to the resulting problem being a mixed-integer problem (MIP) \cite{raman2014model, sadraddini2015robust}.
In order to utilize gradient-based algorithms for faster solution speed, these functions are smoothed using various approximation techniques \cite{pant2017smooth, haghighi2019control, gilpin2020smooth, varnai2020robustness, mao2022successive}.
However, these approximations compromise the soundness or completeness properties of STL, potentially introducing reliability or optimality issues where the solution is either not feasible or not optimal.
Moreover, the use of $\min$ and $\max$ functions may cause locality \& masking problems \cite{mehdipour2019arithmetic}, where the optimizer cannot find a feasible solution even though the problem is feasible. Therefore, the $\min$ and $\max$ functions may not be suitable to use in local optimization algorithms (see \cite{dgmsr} for further discussion).
A generalized mean-based smooth robustness measure for STL specifications (D-GMSR) is proposed in \cite{dgmsr} to alleviate these issues. 
In conjunction with its smoothness, D-GMSR is proven to be both sound and complete. Furthermore, it demonstrates favorable gradient properties essential for numerical optimization purposes and addresses locality \& masking problems.
This paper employs D-GMSR for a smooth, sound, and complete parameterization of the compound state-triggered constraints and uses a real-time implementable CT-SCvx framework for the solution of the resulting PDG problem with continuous-time constraints satisfaction and guaranteed convergence. 

The paper is organized as follows. 
Section \ref{sec:modeling} presents the PDG problem. Section \ref{sec:solution} presents the parameterization of the state-triggered constraints via D-GMSR
and the CT-SCvx framework to solve the resulting PDG problem.
Numerical results are presented in Section \ref{sec:results} and 
the paper is concluded in Section \ref{sec:conclusion}.
\section{6-DoF Powered Descent Guidance Problem \\ with Continuous-time Compound State-Trigerred Constraints} \label{sec:modeling}
This section presents the 6-DoF powered descent guidance problem with continuous-time compound state-triggered constraints. First, we outline the rocket dynamics, general state, control constraints, and boundary conditions from \cite{szmuk2020successive}. 
The gimbaled engine model from \cite{Kamath2023} is incorporated as the control input, leading to a slight adjustment in the dynamics and control constraints. Additionally, the control input includes gimbal deflection and azimuth angles, which define a boresight vector used to impose an altitude-triggered line-of-sight constraint.
We then describe the speed and tilt angle-triggered thrust and altitude-triggered gimbal deflection angle, speed, angular velocity, tilt angle, glideslope, and line-of-sight constraints.

\vspace{-0.45cm}
\subsection{Rocket Dynamics:}
The state vector and the control input of the vehicle are defined as follows:
\begin{align*}
    x :=& \; (m, r_{\mathcal{I}}, v_{\mathcal{I}}, q_{\mathcal{B} \leftarrow \mathcal{I}}, \omega_{\mathcal{B}})\\
    u :=& \; (T, \delta^{\mathrm{e}}, \phi^{\mathrm{e}}, \delta^{\mathrm{b}}, \phi^{\mathrm{b}})
\end{align*}
where subscripts $\mathcal{I}$ and $\mathcal{B}$ are used to denote problem parameters expressed in the inertial and body-fixed reference frames, respectively,
$m \in \mathbb{R}_+$ is the mass of the vehicle, 
$r_{\mathcal{I}} \in \mathbb{R}^3$
is the position vector, 
$v_{\mathcal{I}} \in \mathbb{R}^3$ 
is the velocity vector, 
$q_{\mathcal{B} \leftarrow \mathcal{I}} := (q_1, q_2, q_3, q_4) \in \mathbb{R}^4$ is the quaternion vector that parameterizes the transformation from the inertial frame to the body frame, 
$\omega_{\mathcal{B}} \in \mathbb{R}^3$
is the angular velocity vector, 
$T \in \mathbb{R}$ represents the thrust force, the angles $\delta^{\mathrm{e}}$ and $\phi^{\mathrm{e}}$ correspond to the gimbal deflection and azimuth angles of the engine, as introduced in \cite{Kamath2023}, similarly, $\delta^{\mathrm{b}}$ and $\phi^{\mathrm{b}}$ denote the gimbal deflection and azimuth angles associated with the boresight vector.

The dynamical system of the rocket is given by the following equation:
\begin{align*}
    \Dot{x}(t) &:=  F\big(x(t), u(t)\big)\\ 
    &= 
    \begin{bmatrix}
        - \alpha_{\dot{m}} \| T_{\mathcal{B}} (t) \|_2 \\
        v_{\mathcal{I}} (t) \\
        \frac{1}{m(t)} C_{\mathcal{I} \leftarrow \mathcal{B}}(t) \big( T_{\mathcal{B}}(t) + A_{\mathcal{B}}(t) \big) + g_{\mathcal{I}} \\
        \frac{1}{2} \Omega\big(w_{\mathcal{B}}(t)\big) q_{\mathcal{B} \leftarrow \mathcal{I}} (t) \\
        J_{\mathcal{B}}(t)^{-1} \Big( [r_{\mathrm{cm},\mathcal{B}} \times] T_{\mathcal{B}}(t)
        +
        [r_{\mathrm{cp},\mathcal{B}} \times] A_{\mathcal{B}}(t)
        - \omega_{\mathcal{B}} (t) \times \big(J_{\mathcal{B}}(t) \omega_{\mathcal{B}} \big) \Big)  \nonumber
    \end{bmatrix} 
\end{align*}

In this model, $\alpha_{\dot{m}}$ is given by $\frac{1}{I_{\mathrm{sp}} g_0}$, where $I_{\mathrm{sp}}$ is the vacuum specific impulse of the engine, and $g_0$ is the standard Earth gravity. The gravity vector is denoted as $g_{\mathcal{I}} := (0, 0, -g_0)$.

The moment of inertia matrix of the vehicle about its center of mass at time $t$ is represented by $J_{\mathcal{B}}(t) \in \mathbb{R}_{+}^{3 \times 3}$. The thrust moment-arm, which is the vector from the vehicle's center of mass to the engine gimbal hinge point, is represented by $r_{\mathrm{cm},\mathcal{B}} \in \mathbb{R}^{3}$. The aerodynamic moment-arm, which is the vector from the vehicle's center of mass to the center of pressure, is represented by $r_{\mathrm{cp},\mathcal{B}} \in \mathbb{R}^{3}$.

The thrust vector 
$T_{\mathcal{B}} \in \mathbb{R}^3$
is defined as:
\begin{equation*}
T_{\mathcal{B}} :=
\begin{bmatrix}
    T \sin ( \delta^{\mathrm{e}} ) \cos ( \phi^{\mathrm{e}} ) \\
    T \sin ( \delta^{\mathrm{e}} ) \sin ( \phi^{\mathrm{e}} ) \\
    T \cos ( \delta^{\mathrm{e}} )
\end{bmatrix}
\end{equation*}
where $T$ is the thrust magnitude, $\delta^{\mathrm{e}}$ and $\phi^{\mathrm{e}}$ represent the gimbal deflection and azimuth angles of the engine, respectively.

The aerodynamic force $A_{\mathcal{B}}$ is given by:
\begin{equation*}
A_{\mathcal{B}}(t) = -\frac{1}{2} \rho \|v_{\mathcal{I}}(t)\|_2 S_A C_A C_{\mathcal{B} \leftarrow \mathcal{I}}(t) v_{\mathcal{I}}(t)
\end{equation*}
where $\rho$ is the ambient atmospheric density, $S_A \in \mathbb{R}_{++}$ is the reference area, $C_A \in \mathbb{R}_{++}^{3 \times 3}$ is the aerodynamic coefficient matrix, and $C_{\mathcal{B} \leftarrow \mathcal{I}} = C_{\mathcal{I} \leftarrow \mathcal{B}}^{\top} \in SO(3)$ is the direction cosine matrix that encodes the attitude transformation from the inertial frame to the body frame. The matrix $C_{\mathcal{B} \leftarrow \mathcal{I}}$ is defined as:
\vspace{-0.1cm}
\[
C_{\mathcal{B} \leftarrow \mathcal{I}} =
\begin{bmatrix}
    1 - 2(q_3^2 + q_4^2) & 2(q_2 q_3 + q_1 q_4) & 2(q_2 q_4 - q_1 q_3) \\
    2(q_2 q_3 - q_1 q_4) & 1 - 2(q_2^2 + q_4^2) & 2(q_3 q_4 + q_1 q_2) \\
    2(q_2 q_4 + q_1 q_3) & 2(q_3 q_4 - q_1 q_2) & 1 - 2(q_2^2 + q_3^2)
\end{bmatrix}
\]

The skew-symmetric matrix operator $\Omega(\cdot)$ and the vector cross product $[\cdot \times]$ are defined as:
\begin{equation*}
\Omega(\xi) :=
\begin{bmatrix}
    0 & -\xi_1 & -\xi_2 & -\xi_3 \\
    \xi_1 & 0 & \xi_3 & -\xi_2 \\
    \xi_2 & -\xi_3 & 0 & \xi_1 \\
    \xi_3 & \xi_2 & -\xi_1 & 0
\end{bmatrix}, 
\quad
[\xi \times] :=
\begin{bmatrix}
    0 & -\xi_3 & \xi_2 \\
    \xi_3 & 0 & -\xi_1 \\
    -\xi_2 & \xi_1 & 0
\end{bmatrix}
\end{equation*}
where $\xi := (\xi_1, \xi_2, \xi_3) \in \mathbb{R}^3$.

\vspace{-0.4cm}
\subsection{General state and control constraints}
The vehicle's fuel consumption, tilt angle ($\theta$), angular rate ($\omega_{\mathcal{B}}$), and glideslope angle ($\gamma$) are subject to the following constraints:
\begin{align*} 
    m_{\mathrm{dry}} &\leq m(t)
    \\
    \cos (\theta_{\mathrm{max}}) & \leq 1 - 2 \big( q_{2}^2(t) + q_{3}^2 (t) \big) 
    \\
    \| \omega_{\mathcal{B}}(t) \|_2 &\leq \omega_{\mathrm{max}} 
    \\
    \tan (\gamma_{\mathrm{max}}) \big\| [e_1 \; e_2]^{\top} r_{\mathcal{I}}(t) \big\|_2 &\leq e_3^{\top} r_{\mathcal{I}}(t) 
\end{align*}
where $e_i$ represents the $i$-th basis vector of $\mathbb{R}^3$.

Additionally, the gimbal deflection and azimuth angles of the engine and the boresight vector are constrained by:
\begin{align*} 
    \| \delta^{\mathrm{e}}(t) \|_1 \leq \delta_{\mathrm{max}}^{\mathrm{e}}, \; & \; \| \delta^{\mathrm{b}}(t) \|_1 \leq \delta_{\mathrm{max}}^{\mathrm{b}}\\
    \| \phi^{\mathrm{e}}(t) \|_1 \leq \phi_{\mathrm{max}}^{\mathrm{e}}, \; & \; \| \phi^{\mathrm{b}}(t) \|_1 \leq \phi_{\mathrm{max}}^{\mathrm{b}}
\end{align*}

Since the thrust force is a speed and tilt angle-triggered constraint, it is not explicitly constrained in this section.

For clarity, we represent the state and the control constraints as $g_x(x) \leq 0_{4 \times 1}$ and $g_u(u) \leq 0_{4 \times 1}$, where $g_x : \mathbb{R}^{n_x} \to \mathbb{R}^{4}$ and $g_u : \mathbb{R}^{n_u} \to \mathbb{R}^{4}$. Both $g_x$ and $g_u$ are convex functions, with convexity applied component-wise to each output.
  
\vspace{-0.4cm}
\subsection{Boundary conditions}
The boundary conditions for the problem are defined as follows:
\begin{align*}
    m(0) = m_{\mathrm{i}}, \; m(t_{\mathrm{f}}) \geq m_{\mathrm{dry}}, \; r_{\mathcal{I}}(0) = r_{\mathrm{i}}, \; r_{\mathcal{I}}(t_{\mathrm{f}}) = r_{\mathrm{f}}, \; {v_{\mathcal{I}}(0)} = v_{\mathrm{i}}, \; v_{\mathcal{I}}(t_{\mathrm{f}}) = v_{\mathrm{f}} 
    \\
    q_{\mathcal{B} \leftarrow \mathcal{I}}(0) = q_{{\mathcal{B} \leftarrow \mathcal{I}}_\mathrm{i}}, \; q_{{\mathcal{B} \leftarrow \mathcal{I}}}(t_{\mathrm{f}}) = q_{{\mathcal{B} \leftarrow \mathcal{I}}_\mathrm{f}}, \; \omega_{\mathcal{B}}(0) = \omega_{{\mathcal{B}}_\mathrm{i}}, \; \omega_{\mathcal{B}}(t_{\mathrm{f}}) = \omega_{{\mathcal{B}}_\mathrm{f}} 
\end{align*}
where $m_{\mathrm{dry}}$ is the dry mass of the vehicle, and $i$ and $f$ subscripts denote the initial and final conditions, respectively. For clarity, the boundary conditions are represented as $\big( x(0), x(t_{\mathrm{f}}) \big) \in \mathcal{X}$, where $\mathcal{X}$ is a closed convex set.

\vspace{-0.4cm}
\subsection{Compound state-triggered constraints} \label{sec:stc}
The altitude-triggered gimbal deflection angle, speed, angular velocity, tilt angle and glideslope constraints are defined as follows:
\begin{align*}
    \bigg( 
    e_3^{\top} r_{\mathcal{I}}(t) < h_1^{\mathrm{trig}} \bigg) \implies \bigg( \delta(t) \leq \delta^{\mathrm{stc}} 
    \bigg) 
    & \wedge
    \bigg(
    \| v_{\mathcal{I}}(t) \|_2 \leq v_{\mathcal{I}}^{\mathrm{stc}}
    \bigg) \\
    & \wedge
    \bigg(
    \| \omega_{\mathcal{B}}(t) \|_2 \leq \omega^{\mathrm{stc}}
    \bigg) \\
    & \wedge
    \bigg(
    \cos( \theta^{\mathrm{stc}} ) \leq 1 - 2 \big( q_2^2(t) + q_3^2(t) \big)
    \bigg) \\
    & \wedge
    \bigg(
    \tan (\gamma^{\mathrm{stc}}) \big\| [e_1 \; e_2]^{\top} r_{\mathcal{I}}(t) \big\|_2 \leq e_3^{\top} r_{\mathcal{I}}(t)
    \bigg) 
\end{align*}
where $\implies$ and $\wedge$ denote \textit{implication} and \textit{conjunction} (logical \textit{and}) operators, respectively.


The altitude-triggered line-of-sight angle ($\psi$) constraint is defined as follows:
\begin{align*}
    \bigg( 
    e_3^{\top} r_{\mathcal{I}}(t) < h_2^{\mathrm{trig}} \bigg) \implies \bigg( 
    \cos \big( \psi^{\mathrm{stc}} \big)  \| &r_{\mathcal{I}}(t) - r_{\mathrm{f}} \|_2 \leq \big( r_{\mathcal{I}}(t) - r_{\mathrm{f}} \big)^{\top} C_{\mathcal{I} \leftarrow \mathcal{B}}(t) \ell_{\mathcal{B}}(t) 
    \bigg) 
\end{align*}
where $\psi^{\mathrm{stc}}$ is the view cone angle, and $\ell_{\mathcal{B}}$ is the boresight vector in the body frame, defined as follows:
\begin{align*}
    \ell_{\mathcal{B}}(t) :=
    \begin{bmatrix}
        \sin \big( \delta^{\mathrm{b}} (t) \big) \cos \big( \phi^{\mathrm{b}}(t) \big) \\
        \sin \big( \delta^{\mathrm{b}}(t) \big) \sin \big( \phi^{\mathrm{b}}(t) \big) \\
        \cos \big( \delta^{\mathrm{b}}(t) \big)
    \end{bmatrix}
\end{align*}

The speed and tilt angle triggered thrust force constraints are as follows:
\begin{equation*}
\bigg(
\| v_{\mathcal{I}}(t) \|_2 < v_{\mathcal{I}}^{\mathrm{trig}} \bigg) 
\wedge
\bigg(
\cos( \theta^{\mathrm{trig}} ) < 1 - 2 \big( q_2^2(t) + q_3^2(t) \big)
\bigg)
\implies
\bigg(
T_{\mathrm{min}}^{\mathrm{stc}_1} \leq T(t)
\bigg)
\wedge
\bigg(
T(t) \leq T_{\mathrm{max}}^{\mathrm{stc}_1}
\bigg)
\end{equation*}

If either the speed or the tilt angle of the vehicle exceeds the specified threshold, the thrust force must satisfy the following constraint:
\begin{equation*}
\bigg(
v_{\mathcal{I}}^{\mathrm{trig}} < \| v_{\mathcal{I}}(t) \|_2 \bigg) 
\vee
\bigg(
1 - 2 \big( q_2^2(t) + q_3^2(t) \big) < \cos( \theta^{\mathrm{trig }} )
\bigg)
\implies
\bigg(
T_{\mathrm{min}}^{\mathrm{stc}_2} \leq T(t)
\bigg)
\wedge
\bigg(
T(t) \leq T_{\mathrm{max}}^{\mathrm{stc}_2}
\bigg)
\end{equation*}
where $\vee$ denotes \textit{disjunction} (logical \textit{or}). 
%

For clarity, the state-triggered constraints are represented as:
\begin{subequations} \label{eq:stcs}
\begin{align}
    \big( g^{\mathrm{trig}_1}(x) < 0 \big) 
    & \implies 
    \land_{i=1}^5 \big( g^{\mathrm{stc}_i}(x) \leq 0 \big) \\
    \big( g^{\mathrm{trig}_2}(x) < 0 \big) 
    & \implies 
    \big( g^{\mathrm{stc}_6}(u) \leq 0 \big) \\
    \big( g^{\mathrm{trig}_3}(x) < 0 \big) \wedge \big( g^{\mathrm{trig}_4}(x) < 0 \big) 
    & \implies 
    \big( g^{\mathrm{stc}_7}(u) \leq 0 \big)
    \wedge
    \big( g^{\mathrm{stc}_8}(u) \leq 0 \big)\\
    \big( -g^{\mathrm{trig}_3}(x) < 0 \big) \vee \big( -g^{\mathrm{trig}_4}(x) < 0 \big) 
    & \implies 
    \big( g^{\mathrm{stc}_9}(u) \leq 0 \big)
    \wedge
    \big( g^{\mathrm{stc}_{10}}(u) \leq 0 \big) 
\end{align}
\end{subequations}

\vspace{-0.4cm}
\section{Sequential Convex Programming based Solution Method} \label{sec:solution}
This section presents the parametrization of the compound state-triggered constraints using D-GMSR \cite{dgmsr}, along with the continuous-time successive convexification framework \cite{ctcs2024}, to solve the resulting powered descent guidance problem while ensuring continuous-time constraint satisfaction and guaranteed convergence.

\vspace{-0.2cm}
\subsection{Parametrization of Temporal and Logical Specifications} \label{sec:stl}
Signal Temporal Logic (STL), introduced for monitoring continuous-time signals in \cite{maler2004monitoring}, provides an enriched framework for encoding both temporal and logical specifications, with compound state-triggered constraints being a subset of the logical specifications. Robustness measures in STL quantify how well a signal satisfies a formula and are essential for encoding STL specifications in optimization problems. Several robustness measures have been developed in \cite{donze2010robust, pant2017smooth, gilpin2020smooth, mehdipour2019arithmetic, varnai2020robustness, dgmsr}.

We use the generalized mean-based smooth robustness measure (D-GMSR) proposed in \cite{dgmsr} to parametrize STL specifications. This measure is $\mathcal{C}^1$-smooth, sound, and complete, ensuring that an STL specification is satisfied if and only if its corresponding constraint is satisfied and it enables solving the resulting optimization problem with robust and efficient SCP algorithms while preserving theoretical guarantees. The D-GMSR also resolves issues like locality and masking, which arise from using $\min$ and $\max$ in other STL robustness measures. 
These issues can impact optimization convergence, as discussed in \cite{dgmsr}. A comparison of D-GMSR with other STL measures is summarized in Table \ref{tab:evaluation}, demonstrating that D-GMSR satisfies all the necessary qualities for an accurate and robust representation of STL specifications.
\begin{table}[htbp]
  \centering
  \renewcommand{\arraystretch}{1.0}
  \begin{tabular}
  {|c|c|c|c|c|c|c|}
    \hline
    & \cite{donze2010robust}  & \cite{pant2017smooth} & \cite{gilpin2020smooth} & \cite{mehdipour2019arithmetic} & \cite{varnai2020robustness} & D-GMSR \\
    \hline
    $\mathcal{C}^1$-smoothness          & \redcross  & \greentick      & \greentick       & \redcross   & \redcross   & \greentick \\
    \hline
    Soundness           & \greentick & \yellowcircle   & \greentick       & \greentick  & \greentick      & \greentick \\
    \hline
    Completeness        & \greentick & \yellowcircle   & \yellowcircle    & \greentick  & \greentick      & \greentick \\
    \hline
    Monotonicity        & \greentick & \greentick      & \greentick       & \greentick  & \redcross       & \greentick \\
    \hline
    Locality \& Masking & \redcross  & \purpletriangle & \purpletriangle  & \greentick  & \greentick & \greentick \\
    \hline
  \end{tabular}
  \caption{\small Comparison of the D-GMSR with the previous robustness measures \\ (\yellowcircle: Satisfied only for very large smoothing parameters;
  \purpletriangle: Satisfied only for small smoothing parameters)}
  \label{tab:evaluation}
  \vspace{-14pt}
\end{table}

We begin by summarizing the syntax of the STL and then introduce D-GMSR for parameterizing the compound state-triggered constraints in the powered descent guidance.

\subsubsection{STL syntax}
STL syntax is recursively defined for a non-empty time interval $ I = \intv{a}{b} := \{ a, a+1, \dots, b \} $ as follows:  
\begin{equation*} \label{eq:stl_syntax}
    \varphi ::= \mu \; | \; \neg \varphi_1 \; | \; \varphi_1 \wedge \varphi_2 \; | \; \varphi_1 \bm{U}_I \varphi_2
\end{equation*}
Here, $\varphi$ denotes an STL formula, and $\varphi_1, \varphi_2$ are sub-formulas. The operators $\neg$ (\textit{negation}), $\wedge$ (\textit{conjunction}), and $\bm{U}_I$ (\textit{until}) combine predicates or sub-formulas to construct STL specifications. Table \ref{tab:stl_opt} illustrates how additional Boolean and temporal operators, such as \textit{disjunction} ($\vee$), \textit{implication} ($\implies$), \textit{eventually} ($\bm{F}_I$), and \textit{always} ($\bm{G}_I$), are derived.
\begin{table}[H]
    \centering
    \renewcommand{\arraystretch}{1.1}
    \begin{tabular}{|cc|c|}
    \hline
    \multicolumn{2}{|c|}{STL Specification}     & Formula \\ \hline
    \multicolumn{1}{|c|}{\textit{Disjunction}} & $\varphi_1 \vee \varphi_2$ & $\neg (\neg \varphi_1 \wedge \neg \varphi_2)$ \\ \hline
    \multicolumn{1}{|c|}{\textit{Implication}} & $\varphi_1 \implies \varphi_2$ & $\neg \varphi_1 \vee \varphi_2$ \\ \hline
    \multicolumn{1}{|c|}{\textit{Eventually}} & $\bm{F}_I \varphi$ & $\top \bm{U}_I \varphi$ \\ \hline
    \multicolumn{1}{|c|}{\textit{Always}} & $\bm{G}_I \varphi$ & $\neg \bm{F}_I \neg \varphi$ \\ \hline
    \end{tabular}
    \caption{Derived Boolean and Temporal Operators in STL}
    \label{tab:stl_opt}
\end{table}
\vspace{-0.5cm}
\begin{rem}
The $\bm{U}_I$ operator — and consequently the $\bm{F}_I$ and $\bm{G}_I$ operators — can be expressed in discrete time using only the predicate, \textit{negation}, and \textit{conjunction}. Therefore, \textit{conjunction} and \textit{disjunction} operators are sufficient for defining discrete-time STL specifications (see \cite{dgmsr} for details).
\end{rem} 

\vspace{-0.5cm}
\subsubsection{Generalized mean-based smooth robustness (D-GMSR)}
To parametrize the \textit{conjunction} and \textit{disjunction} operators, the functions ${}^{\wedge} h_{p, w}^{\shft}$ and ${}^{\vee} h_{p, w}^{\shft}$ are constructed using generalized means, such as the weighted geometric mean and weighted power mean, as follows:

\begin{equation*}
\begin{aligned}
    {}^{\wedge} h_{p, w}^{\shft}(y) &:= 
    \bigg( M_{0, w}^{\shft}(|y|_{+}^2) \bigg)^{\frac{1}{2}}
    - \bigg( M_{p, w}^{\shft}(|y|_{-}^2) \bigg)^{\frac{1}{2}} \\
    {}^{\vee} h_{p, w}^{\shft}(y) &:= -{}^{\wedge} h_{p, w}^{\shft}(-y)
\end{aligned}
\end{equation*}
where
\vspace{-0.25cm}
\begin{center}
\begin{minipage}{0.4\textwidth}
    \begin{align*} 
        M_{0, w}^{\shft}(z) &:= \Big( \shft^{\bm{1}^{\top} w} + \prod_{i=1}^n z_i^{w_i} \Big)^{1/\bm{1}^{\top} w}
    \end{align*}
\end{minipage}%
\hspace{0.25cm}
\begin{minipage}{0.4\textwidth}
    \begin{align*} 
        M_{p, w}^{\shft}(z) &:= \Big( \shft^p + \frac{1}{\bm{1}^{\top} w} \sum_{i=1}^n w_i z_i^p \Big)^{1/p}
    \end{align*}
\end{minipage}%
\end{center}
where $y \in \mathbb{R}^n$, 
$\shft \in \mathbb{R}_{++}$, $p \in \mathbb{Z}_{++}$, $w \in \mathbb{Z}_{++}^n$. Hence, $(f_i(x) \geq 0\big)$, $i=1,2,\dots,n$ specifications, composed via \textit{conjunction} and \textit{disjunction} operators, are parametrized using the ${}^{\wedge} h_{p, w}^{\shft}$ and ${}^{\vee} h_{p, w}^{\shft}$ functions as follows:
\begin{align*} 
    \land_{i=1}^n \big(f_i(x) \geq 0\big) &\iff {}^{\wedge} h_{p, w}^{\shft}  \Big( \big( f_1(x), f_2(x), \dots, f_n(x) \big) \Big) \geq 0  \\
    \lor_{i=1}^n  \big(f_i(x) \geq 0\big) &\iff {}^{\vee} h_{p, w}^{\shft}  \Big( \big( f_1(x), f_2(x), \dots, f_n(x) \big) \Big) \geq 0
\end{align*}

\begin{rem}
To ensure the satisfaction of a specification parametrized with the key functions of the D-GMSR, ${}^{\wedge} h_{p, w}^{\shft}$ and  ${}^{\vee} h_{p, w}^{\shft}$, can be simplified as follows:
\vspace{-0.25cm}
\begin{center}
\begin{minipage}{0.4\textwidth}
    \begin{align*}
        {}^{\wedge} h_{p, w}^{\shft}( y ) &\geq 0 \\
        \iff \shft^{0.5} - (M_{p,w}^{\shft}(|y|_{-}^2))^{0.5} &= 0 \\
        \iff \shft^{0.5} - \bigg( \shft^p + \frac{1}{\bm{1}^{\top} w} \sum_{i=1}^n w_i\min(0, y_i)^{2p}  \bigg)^{\frac{1}{2p}} &= 0 \\
        \iff  \sum_{i=1}^n \max(0, y_i)^{2}  &= 0
    \end{align*}
\end{minipage}%
\hfill
\begin{minipage}{0.4\textwidth}
    \begin{align*}
        {}^{\vee} h_{p, w}^{\shft}( y ) &\geq 0 \\
        \iff \shft^{0.5} - (M_{0,w}^{\shft}(|y|_{-}^2))^{0.5} &= 0 \\
        \iff \shft^{0.5} - \bigg( \shft^{\bm{1}^{\top} w} + \prod_{i=1}^n \min(0, y_i)^{2w_i}  \bigg)^{\frac{0.5}{\bm{1}^{\top} w}} &= 0 \\
        \iff \prod_{i=1}^n \max(0, y_i)^{2}   &= 0
    \end{align*}
\end{minipage}%
\end{center}
Note that, although these equalities are mathematically equivalent, they can lead to varying numerical optimization performance, making the choice of formulation crucial for problem design.
\end{rem}

This simplification can be generalized for an arbitrary number of combinations of \textit{conjunction} and \textit{disjunction} operators. For example, the satisfaction of the specification:
$$\Big( \big( f_1(x) \geq 0 \big) \wedge \big( f_2(x) \geq 0 \big) \Big) \vee \Big( \big( f_3(x) \geq 0 \big) \wedge \big( f_4(x) \geq 0 \big) \Big) $$
can be ensured by the satisfaction of the following constraints:
\vspace{-0.2cm}
\begin{align*}
    {}^{\vee} h_{1, \bm{1}}^{\shft}
    \bigg( 
    \bigg(
    {}^{\wedge} h_{1, \bm{1}}^{\shft}\Big( \big(f_1(x), f_2(x)\big) \Big),
    {}^{\wedge} h_{1, \bm{1}}^{\shft}\Big( \big(f_3(x), f_4(x)\big) \Big)
    \bigg)
    \bigg) &\geq 0 \\
    \iff \Big( \max(0, f_1(x))^2 + \max(0, f_2(x))^2 \Big) \Big( \max(0, f_3(x))^2 + \max(0, f_4(x))^2 \Big) &= 0
\end{align*}

In this regard, STCs in Eq. (\ref{eq:stcs}) can be parametrized as follows:
\vspace{-0.25cm}
\begin{subequations} \label{eq:stc_param}
    \begin{align}
        \max\big(0, - g^{\mathrm{trig}_1}(x)  \big)^2 
        \bigg( \sum_{i=1}^5 \max\big(0, g^{\mathrm{stc}_i}(x)\big)^2 \bigg)
        &= 0 \\
        \Bigg(
        \max\big(0, - g^{\mathrm{trig}_2}(x)  \big)^2 
        \max\big(0, g^{\mathrm{stc}_6}(u)\big)^2
        \Bigg)  &= 0 \\
        \! \bigg(
        \max\big(0, - g^{\mathrm{trig}_3}(x)  \big)^2 
        \max\big(0, - g^{\mathrm{trig}_4}(x)  \big)^2 
        \Big( \max\big(0, g^{\mathrm{stc}_7}(u)\big)^2 + \max\big(0, g^{\mathrm{stc}_8}(u)\big)^2 \Big)
        \bigg) &= 0 \\
        \bigg(
        \Big(
        \max\big(0,  g^{\mathrm{trig}_3}(x)  \big)^2 
        +
        \max\big(0,  g^{\mathrm{trig}_4}(x)  \big)^2 
        \Big) 
        \Big( \max\big(0, g^{\mathrm{stc}_9}(u)\big)^2 + \max\big(0, g^{\mathrm{stc}_{10}}(u)\big)^2 \Big)
        \bigg) &= 0
    \end{align}
\end{subequations}
\begin{rem}
    Note that, all the functions in Eq. (\ref{eq:stc_param}) are nonnegative, which eliminates the need to use exterior penalty functions to augment the dynamics to ensure continuous-time constraint satisfaction (see \cite[Section 2.3]{ctcs2024}).
\end{rem}

\vspace{-0.15cm}
For clarity, we represent the state-triggered constraints in Eq. \eqref{eq:stc_param} as:
\[
h_{\mathrm{stc}}\big(x(t), u(t)\big) = 0_{4 \times 1}, \quad \text{where} \quad h_{\mathrm{stc}} : \mathbb{R}^{n_x \times n_u} \to \mathbb{R}_+^4
\]

The resulting free-final-time powered descent guidance problem is presented as follows:
\begin{eqbox}
\hspace{0.2cm} \textbf{
The 
free-final-time 
infinite-dimensional 
nonconvex 
optimal control problem
}
\begin{subequations} \label{ocp}
\begin{align} 
\underset{t_{\mathrm{f}},\,x,\,u}{\operatorname{minimize}}~\,&~ t_{\mathrm{f}} \label{ocp:obj} \\ %
\operatorname{subject~to}\,
    &~\dot{x}\big(t\big) = F\big(x(t),u(t)\big), \quad\;\;\,\, t \in [0, t_{\mathrm{f}}] \label{ocp:dyn}\\
    &~g_x\big(x(t)\big) \leq 0_{4 \times 1}, \qquad\quad\;\, t \in [0, t_{\mathrm{f}}] \label{ocp:path_x}\\
    &~g_u\big(u(t)\big) \leq 0_{4 \times 1}, \qquad\quad\;\, t \in [0, t_{\mathrm{f}}] \label{ocp:path_u}\\
    &~h_{\mathrm{stc}}\big(x(t), u(t)\big) = 0_{4 \times 1}, \quad t \in [0, t_{\mathrm{f}}] \label{ocp:stc_1}\\
    &~\big( x(0), x(t_{\mathrm{f}}) \big) \in \mathcal{X} \label{ocp:bnd}
\end{align}%
\end{subequations}
\end{eqbox}
\noindent

\vspace{-0.25cm}
\subsection{Continuous-time Successive Convexification Framework:}
This section presents the continuous-time successive convexification framework \cite{ctcs2024} for the solution of the powered descent guidance problem with continuous-time state-triggered constraints.

\vspace{-0.3cm}
\subsubsection{Reformulation of the path constraints}
To ensure the satisfaction of the path constraints in continuous-time, the constraints are reformulated as follows:
\begin{equation*}
    \begin{aligned}
    & 
    g_x\big(x(t)\big) \leq 0_{4 \times 1}, \;
    g_u\big(u(t)\big) \leq 0_{4 \times 1}, \;
    h_{\mathrm{stc}}\big(x(t), u(t)\big) = 0_{4 \times 1}, \;
    \;\; \mathrm{ a. e. } \;  t \in [0, t_{\mathrm{f}}] \\
    & \iff
    \int_{0}^{t_{\mathrm{f}}} \!\!\! 
    1_{4 \times 1}^{\top} q_c( 0, g_x\big(x(t)\big) ) +
    1_{4 \times 1}^{\top} q_c( 0, g_u\big(u(t)\big) ) +
    1_{4 \times 1}^{\top} h_{\mathrm{stc}}\big(x(t), u(t)\big) \mathrm{d} t = 0 \\
    & \iff
    \dot{y}(t) = 
    1_{4 \times 1}^{\top} q_c( 0, g_x\big(x(t)\big) ) + 1_{4 \times 1}^{\top} q_c( 0, g_u\big(u(t)\big) ) + 
    1_{4 \times 1}^{\top} h_{\mathrm{stc}}\big(x(t), u(t)\big), \text{\; and \;} y(0) = y(t_{\mathrm{f}})
    \end{aligned}
\end{equation*}
where $q_c(x) = \max(0, x)^2 $ for $q: \mathbb{R}^n \to \mathbb{R}^n$ and $c \in R_+$. 

The system dynamics are then augmented with a new state variable $y \in \mathbb{R}_+$, yielding:
\begin{align*}
    \dot{x}_{\mathrm{a}}(t) & = \left[ \begin{array}{c} \dot{x}(t) \\[0.05cm]  \dot{y}(t) \end{array} \right]\nonumber\\[0.35cm]
    & = \left[ \begin{array}{c} F(x(t),u(t)) \\[0.05cm]  
    1_{4 \times 1}^{\top} q_c( 0, g_x\big(x(t)\big) ) +
    1_{4 \times 1}^{\top} q_c( 0, g_u\big(u(t)\big) ) +
    1_{4 \times 1}^{\top} h_{\mathrm{stc}}\big(x(t), u(t)\big) \end{array} \right]\\[0.35cm]
    & = F_{\mathrm{a}}(x_{\mathrm{a}}(t),u(t))
\end{align*}

We employ the exterior penalty function $q_c$ to reformulate the path constraints in Eq. (\ref{ocp:path_x}) and (\ref{ocp:path_u}); however, the path constraints in Eq. (\ref{ocp:stc_1}) does not require an exterior penalty function because it is inherently a nonnegative function. 
Additionally, the need for reformulating the control constraints in Eq. (\ref{ocp:path_u}) depends on the parametrization techniques applied to the control inputs. In some cases, ensuring the satisfaction of these constraints at a finite number of discrete points may be sufficient to guarantee their satisfaction in continuous-time (e.g. first-order-hold parametrization).

The reformulated free-final-time powered descent guidance problem is presented as follows:
\begin{eqbox}
\hspace{0.2cm} \textbf{
The 
reformulated
free-final-time 
infinite-dimensional 
nonconvex 
optimal control problem
}
\begin{subequations} \label{ctcs_ocp}
\begin{align} 
\underset{t_{\mathrm{f}},\,x_{\mathrm{a}},\,u}{\operatorname{minimize}}~\,&~ t_{\mathrm{f}} \label{ctcs_ocp:obj} \\ %
\operatorname{subject~to}\,
    &~\dot{x}_{\mathrm{a}}(t) = F_{\mathrm{a}}(x_{\mathrm{a}}(t),u(t)), \quad\;\, t \in [0, t_{\mathrm{f}}]\label{ctcs_ocp:dyn}\\
    &~{}^{y} E \big(x_{\mathrm{a}}(1)-x_{\mathrm{a}}(0)\big) = 0\label{ctcs_ocp:ctcs}\\
    &~\big( {}^{x} E x_{\mathrm{a}}(0), {}^{x} E x_{\mathrm{a}}(t_{\mathrm{f}}) \big) \in \mathcal{X}
    \label{ctcs_ocp:bnd}
\end{align}%
\end{subequations}
\end{eqbox}
\noindent
where ${}^{x} E$, ${}^{y} E$ select the elements of the $x_{\mathrm{a}}$ corresponding to $x$ and $y$, respectively. Eq. (\ref{ctcs_ocp:ctcs}) ensures the continuous-time satisfaction of the path constraints.

\subsubsection{Time-dilation}
This section presents time-dilation for transforming the free-final-time powered descend guidance problem to a fixed-final-time problem.

Consider a function $\tilde{t}$ that maps a normalized interval to physical time, defined as $\tilde{t} : [0, 1] \to [0, t_\mathrm{f}]$. The rate of change of time is represented by the function $s$ and is expressed as:  
\begin{equation*}
    s(\tau) := \frac{\mathrm{d} \tilde{t}(\tau)}{\mathrm{d}\tau} = \derv{\tilde{t}}(\tau), \quad \tau \in [0,1],
\end{equation*}  
where $\derv{\square}$ denotes the derivative with respect to $\tau$.
The function $s$ is treated as a variable that controls the rate of time progression. By incorporating $s$, the state, control inputs, and system dynamics are redefined as functions of $\tau$, as follows:  
\begin{align*}
    \tilde{x}(\tau) &:= \big( x_{\mathrm{a}}(\tilde{t}(\tau)), \tilde{t}(\tau) \big) \\[0.25cm]
    \tilde{u}(\tau) &:= \big( u(\tilde{t}(\tau)), s(\tau) \big) \\[0.25cm]
    \derv{\tilde{x}}(\tau) &= 
    \begin{bmatrix}
        \dot{x}_{\mathrm{a}}(t) \\
        1
    \end{bmatrix}
    \frac{\mathrm{d} \tilde{t}(\tau)}{\mathrm{d} \tau} \\[0.25cm]
    &= 
    \begin{bmatrix}
        F_{\mathrm{a}}\big( x_{\mathrm{a}}(\tilde{t}(\tau)), u(\tilde{t}(\tau)) \big)\\
        1
    \end{bmatrix}
    s(\tau) \\[0.25cm]
    &= f\big(\tilde{x}(\tau), \tilde{u}(\tau)\big)
\end{align*}

The resulting fixed-final-time powered descent guidance problem is presented as follows:
\begin{eqbox}
\hspace{0.2cm} \textbf{
The 
reformulated
fixed-final-time 
infinite-dimensional 
nonconvex 
optimal control problem
}
\begin{subequations} \label{td_ocp}
\begin{align} 
\underset{\tilde{x},\,\tilde{u}}{\operatorname{minimize}}~\,&~ 
{}^{\tilde{t}} E \tilde{x}(1) \label{td_ocp:obj} \\ %
\operatorname{subject~to}\,
    &~\derv{\tilde{x}}(\tau) = f\big(\tilde{x}(\tau), \tilde{u}(\tau)\big), \quad \tau \in [0, 1]\label{td_ocp:dyn}\\
    &~{}^{y} E (\tilde{x}(1)-\tilde{x}(0)) = 0\label{td_ocp:ctcs}\\
    &~{}^{s} E \tilde{u}(\tau) \geq 0\label{td_ocp:pos_time}\\
    &~\big( {}^{x} E \tilde{x}(0), {}^{x} E \tilde{x}(1) \big) \in \mathcal{X}
    \label{td_ocp:bnd}
\end{align}%
\end{subequations}
\end{eqbox}
\noindent
where ${}^{x} E$, ${}^{y} E$ and ${}^{\tilde{t}} E$ select the elements of the $\tilde{x}$ corresponding to $x$, $y$ and $\tilde{t}$, respectively, ${}^{s} E$ selects the element of the $\tilde{u}$ corresponding to $s$. Eq. (\ref{td_ocp:pos_time}) ensures the non-negativity of the time-dilation factor.

\vspace{-0.3cm}
\subsubsection{Parameterization of the control input and exact discretization of the dynamics} \label{sec:scvx:param}
We first parametrize the control input to transform the infinite-dimensional powered descent guidance problem to a finite-dimensional problem and then discretize the dynamics via multiple-shooting \cite{bock1984multiple, quirynen2015multiple} for numerical tractability.

Suppose the time is discretized over a finite grid as $0 = \tau_1 < \tau_2 < \dots < \tau_K = 1$, $\tilde{x}_k$ and $\tilde{u}_k$ represent the state and control input at the node point $\tau_k$.
The control input is parameterized via first-order hold as
\begin{align*}
    \tilde{u}(\tau) = \bigg( \frac{\tau_{k+1} - \tau}{\tau_{k+1} - \tau_k} \bigg) \tilde{u}_k + \bigg( \frac{\tau - \tau_{k}}{\tau_{k+1} - \tau_k} \bigg) \tilde{u}_{k+1}
\end{align*}
$\forall \tau \in [0,1]$ and $\forall k \in \intv{1}{K-1}$. 

Define 
\vspace{-0.1cm}
\begin{align} \label{dt-dyn}
    F_k(\tilde{x}_k, \tilde{u}_k, \tilde{u}_{k+1}) := \tilde{x}_k + \int_{\tau_k}^{\tau_{k+1}} f( \tilde{x}^k(\tau), \tilde{u}(\tau) ) \mathrm{d}\tau
\end{align}
$\forall k \in \intv{1}{K-1} $, where the state trajectory $\tilde{x}^k$ satisfies Eq. (\ref{td_ocp:dyn}) on $[\tau_k, \tau_{k+1}]$ with initial condition $\tilde{x}_k$ and control input $\tilde{u}$. Then, Eq. (\ref{td_ocp:dyn}) is equivalent to
\begin{align*}
    \tilde{x}_{k+1} - F_k(\tilde{x}_k, \tilde{u}_k, \tilde{u}_{k+1}) = 0
\end{align*}
on $[\tau_k, \tau_{k+1}]$, $\forall k \in \intv{1}{K-1}$.

The equality constraint defined for CTCS in Eq. (\ref{td_ocp:ctcs}) is equivalent to ${}^{y} E (\tilde{x}_{k+1}-\tilde{x}_k) = 0$; however, the constraint is relaxed as ${}^{y} E (\tilde{x}_{k+1}-\tilde{x}_k) 
\leq \epsilon_{\mathrm{LICQ}}$, so that solutions do not violate the linear independence
constraint qualification (LICQ) \cite{ctcs2024}. For each relaxation parameter $\epsilon_{\mathrm{LICQ}} > 0$, there exists a corresponding constraint tightening value $\delta_{\mathrm{LICQ}} > 0$ 
that ensures the satisfaction of inequality constraints by reducing their upper bounds by $\delta_{\mathrm{LICQ}}$ (see \cite{ctcs2024} for details). We define $g^{\mathrm{trig}_i}$, $i \in [1\!:\!4]$ functions in Eq. (\ref{eq:stcs}) accordingly.
The inequality constraint ensuring the nonnegativity of the time-dilation factor in Eq. (\ref{td_ocp:pos_time}) is expressed as ${}^{s} E (\tilde{u}_{k+1} - \tilde{u}_k) \geq 0$. However, to prevent degenerate solutions, this constraint is tightened to ${}^{s} E (\tilde{u}_{k+1} - \tilde{u}_k) \geq s_{\mathrm{min}}$, where $s_{\mathrm{min}} \in \mathbb{R}_{++}$.

The resulting finite-dimensional powered descent guidance problem is presented as follows:
\begin{eqbox}
\hspace{0.2cm} \textbf{
The 
reformulated
fixed-final-time 
finite-dimensional 
nonconvex 
optimal control problem
}
\begin{subequations} \label{fd}
\begin{align} 
\underset{\tilde{x}_k,\,\tilde{u}_k}{\operatorname{minimize}}~\,&~ \frac{1}{2} \bigg( \sum_{i=1}^{K-1} {}^{s} E \tilde{u}_k + \sum_{i=2}^{K} {}^{s} E \tilde{u}_k 
\bigg) \label{fd:obj} \\ %
\operatorname{subject~to}\,
    &~\tilde{x}_{k+1} - F_k(\tilde{x}_k, \tilde{u}_k, \tilde{u}_{k+1}) = 0\label{fd:dyn}\\
    &~{}^{y} E (\tilde{x}_{k+1}-\tilde{x}_k) \leq \epsilon_{\mathrm{LICQ}}\label{fd:ctcs}\\
    &~{}^{s} E (\tilde{u}_{k+1}-\tilde{u}_k) \geq s_{\mathrm{min}} \label{fd:pos_time}\\
    &~\big( {}^{x} E \tilde{x}_1, {}^{x} E \tilde{x}_K \big) \in \mathcal{X}
    \label{fd:bnd}
\end{align}%
\end{subequations}
\end{eqbox}
\noindent
where ${}^{x} E$ and ${}^{y} E$ select the elements of the $\tilde{x}_k$ corresponding to $x$ and $y$, respectively, ${}^{s} E$ selects the element of the $\tilde{u}_k$ corresponding to $s$. Note that, since 
control inputs, including the dilation-factor $s$, is parametrized via first-order-hold ${}^{\tilde{t}} E \tilde{x}(1) = \frac{1}{2} \Big( \sum_{i=1}^{K-1} {}^{s} E \tilde{u}_k + \sum_{i=2}^{K} {}^{s} E \tilde{u}_k 
\Big)$.

\subsubsection{Penalization of the nonconvex constraints}
To solve the resulting finite-dimensional powered descent guidance problem \eqref{fd}, we use an exact penalization method \cite[Chap. 17]{nocedal2006numerical} for the nonconvex constraints. Since the objective function and all the constraints except the dynamic constraint are linear, dynamic constraints are penalized via $L_1$-norm as follows:
\begin{eqbox}
\hspace{0.2cm} \textbf{
The 
reformulated
fixed-final-time 
finite-dimensional 
penalized
nonconvex 
optimal control problem
}
\begin{subequations} \label{pen_fd}
\begin{align} 
\underset{\tilde{x}_k,\,\tilde{u}_k}{\operatorname{minimize}}~\,
&~ \frac{1}{2} \bigg( \sum_{i=1}^{K-1} {}^{s} E \tilde{u}_k + \sum_{i=2}^{K} {}^{s} E \tilde{u}_k 
\bigg) \label{pen_fd:cvx_obj} \\
&~ + w_{\mathrm{eq}}^{\mathrm{dyn}} \sum_{i=1}^{K-1} \| \tilde{x}_{k+1} - F_k(\tilde{x}_k, \tilde{u}_k, \tilde{u}_{k+1}) \|_1 \label{pen_fd:noncvx_obj} \\ %
\operatorname{subject~to}\,
    &~{}^{y} E (\tilde{x}_{k+1}-\tilde{x}_k) \leq \epsilon_{\mathrm{LICQ}} \label{pen_fd:ctcs}\\
    &~{}^{s} E (\tilde{u}_{k+1}-\tilde{u}_k) \geq s_{\mathrm{min}}\label{pen_fd:pos_time}\\
    &~\big( {}^{x} E \tilde{x}_1, {}^{x} E \tilde{x}_K \big) \in \mathcal{X}
    \label{pen_fd:bnd}
\end{align}%
\end{subequations}
\end{eqbox}
\noindent
For a large enough, finite $w_{\mathrm{eq}}^{\mathrm{dyn}}$, we can show that a solution of \eqref{pen_fd} that is feasible is also a solution to \eqref{fd}. See \cite[Chap. 17]{nocedal2006numerical} for a detailed discussion.

\subsubsection{Prox-linear method with adaptive weight update algorithm}

The resulting finite-dimensional nonconvex optimization problem is solved using a convergence-guaranteed SCP algorithm, the prox-linear method \cite{drusvyatskiy2019efficiency}, along with an adaptive weight update algorithm \cite[Algorithm 2.2]{cartis2011evaluation}.

The penalized problem \eqref{pen_fd} can be compactly expressed as:
\begin{align}
    \underset{Z\in\mathcal{Z}}{\mathrm{minimize}}~~J_{\mathrm{nl}}(Z) := G(Z) + w_{\mathrm{ep}} H(C(Z))\label{dt-penal-ocp-compact}
\end{align}
where $Z = (\tilde{x}_1,\ldots,\tilde{x}_K,\tilde{u}_1,\ldots,\tilde{u}_K)$, $\mathcal{Z}$ is a convex set corresponding to constraints in Eq. \eqref{pen_fd:ctcs}-\eqref{pen_fd:bnd}, $G$ is a proper convex function corresponding to \eqref{pen_fd:cvx_obj}, and $H \circ C$ is a composition of a proper convex function and $\mathcal{C}^1$-smooth mapping corresponding to \eqref{pen_fd:noncvx_obj}.

The prox-linear method determines a stationary point of \eqref{dt-penal-ocp-compact} by solving a series of convex subproblems. At iteration \( j+1 \), the following convex problem is solved:
\begin{align} \label{cvx-subproblem}
    \underset{Z\in\mathcal{Z}}{\mathrm{minimize}}~
    J_{\mathrm{lin}}(Z, Z^{j}, w_{\mathrm{prox}}) &:= G(Z) + w_{\mathrm{ep}} H(C(Z^{j}) + \nabla C(Z^{j})(Z-Z^{j})) + \frac{w_{\mathrm{prox}}}{2}\|Z-Z^j\|_2^2
\end{align}
where $Z^j$ is the solution from iteration $j$. The sequence $Z^j$ converges for a suitably chosen weight $w_{\mathrm{prox}}$. Moreover, if the converged point satisfies the feasibility conditions for \eqref{dt-penal-ocp-compact}, it will also be a KKT point.

The specialization of \eqref{cvx-subproblem} at iteration $j+1$ to \eqref{pen_fd} can be stated as:
\begin{eqbox}
\hspace{0.2cm} \textbf{
The 
reformulated
fixed-final-time 
finite-dimensional 
penalized
convex 
optimal control problem
}
\begin{subequations} \label{lpen_fd}
\begin{align} 
\underset{\tilde{x}_k,\,\tilde{u}_k}{\operatorname{minimize}}~\,
&~ \frac{1}{2} \bigg( \sum_{i=1}^{K-1} {}^{s} E \tilde{u}_k + \sum_{i=2}^{K} {}^{s} E \tilde{u}_k 
\bigg) \label{lpen_fd:cvx_obj} \\
~& + w_{\mathrm{eq}}^{\mathrm{dyn}} \sum_{k=1}^{K-1} \| \tilde{x}_{k+1} - (A_k \tilde{x}_k + B^-_k \tilde{u}_k + B^+_k \tilde{u}_{k+1} + w_k) \|_1 \\
&~+ \frac{w_{\mathrm{prox}}}{2} \sum_{k=1}^K \|\tilde{x}_k-\tilde{x}^j_k\|_2^2 + \|\tilde{u}_k -\tilde{u}^j_k\|_2^2\label{lpen_fd:noncvx_obj} \\[-0.05cm] %
\operatorname{subject~to}\,
    &~{}^{y} E (\tilde{x}_{k+1}-\tilde{x}_k) \leq \epsilon_{\mathrm{LICQ}} \label{lpen_fd:ctcs}\\
    &~{}^{s} E (\tilde{u}_{k+1}-\tilde{u}_k) \geq s_{\mathrm{min}}\label{lpen_fd:pos_time}\\
    &~\big( {}^{x} E \tilde{x}_1, {}^{x} E \tilde{x}_K \big) \in \mathcal{X}
    \label{lpen_fd:bnd}
\end{align}%
\end{subequations}
\end{eqbox}
\noindent
where $A_k$, $B^-_k$, $B^+_k$, and $w_k$, $\forall k\in\intv{1}{K-1}$, form the gradient of the dynamical constraints in Eq. \eqref{fd:dyn} at $Z^j = (\tilde{x}^j_1,\ldots,\tilde{x}^j_K,\tilde{u}^j_1,\ldots,\tilde{u}^j_K)$. Specifically, for each $k\in\intv{1}{K-1}$, let $\hat{x}^{j}_k(\tau)$ denote the solution to the dynamics in Eq. \eqref{td_ocp:dyn} over $[\tau_k,\tau_{k+1}]$, which is generated with the initial condition $\tilde{x}_k^j$ and the control input $\tilde{u}^j(\tau)$, parameterized using a first-order hold with $\tilde{u}_k^j$ and $\tilde{u}_{k+1}^j$.
The Jacobians of \( f \) in Eq. \eqref{td_ocp:dyn}, evaluated at \( (\hat{x}_k^j(\tau), \tilde{u}^j(\tau)) \), are given by:
\begin{align*}
    A(\tau) ={} & \left.\frac{\partial f(\tilde{x},\tilde{u})}{\partial \tilde{x}}\right|_{\big(\hat{x}^{j}_k(\tau),\tilde{u}^j(\tau)\big)}\\
    B(\tau) ={} & \left.\frac{\partial f(\tilde{x},\tilde{u})}{\partial \tilde{u}}\right|_{\big(\hat{x}^{j}_k(\tau),\tilde{u}^j(\tau)\big)}
\end{align*}
We then solve the following system of initial value problems:
\begin{align*}
    \dot{\Phi}_{\tilde{x}}(\tau,\tau_k) ={} & A(\tau)\Phi_{\tilde{x}}(\tau,\tau_k)\\
    \dot{\Phi}^{-}_{\tilde{u}}(\tau,\tau_k) ={} & A(\tau)\Phi_{\tilde{u}}^{-}(\tau,\tau_k) + B(\tau)\left(\frac{\tau_{k+1}-\tau}{\tau_{k+1}-\tau_k}\right)\\
    \dot{\Phi}^{+}_{\tilde{u}}(\tau,\tau_k) ={} & A(\tau)\Phi_{\tilde{u}}^{+}(\tau,\tau_k) + B(\tau)\left(\frac{\tau-\tau_k}{\tau_{k+1}-\tau_k}\right)
\end{align*}
\vspace{-0.05cm}
with the initial conditions:
\vspace{-0.05cm}
\begin{align*}
    \Phi_{\tilde{x}}(\tau_k,\tau_k) ={} & I_{n_x}\\
    \Phi_{\tilde{u}}^{-}(\tau_k,\tau_k) ={} & 0_{n_x\times n_u}\\
    \Phi_{\tilde{u}}^{+}(\tau_k,\tau_k) ={} & 0_{n_x\times n_u}
\end{align*}
\vspace{-0.05cm}
Solving this system yields the following values, which form the gradient of the dynamical constraints in Eq. \eqref{fd:dyn}:
\vspace{-0.05cm}
\begin{align*}
    A_k ={} & \Phi_{\tilde{x}}(\tau_{k+1},\tau_k)\\
    B_k^{-} ={} & \Phi_{\tilde{u}}^{-}(\tau_{k+1},\tau_k)\\
    B_k^{+} ={} & \Phi_{\tilde{u}}^{+}(\tau_{k+1},\tau_k)\\
    w_k ={} & \hat{x}^{j}_{k}(\tau_{k+1}) - A_k \tilde{x}^j_k - B^-_k \tilde{u}^j_k - B_k^{+}\tilde{u}^{j}_{k+1} 
\end{align*}

We dynamically determine the proximal term weight in Eq. \eqref{lpen_fd:noncvx_obj} using the following adaptive weight update algorithm presented in \cite[Algorithm 2.2]{cartis2011evaluation}.
\begin{algorithm}[!htpb] 
\caption{Adaptive weight update algorithm}
\label{alg:adaptive_prox}
\begin{algorithmic}[0] 
\State \textbf{Inputs:} 
$w_{\mathrm{prox}}$, $Z^{j}$, $Z^{j+1}$, 
$0 < \beta_1 < \beta_2 $,
$\sigma_3 < 1 < \sigma_2 < \sigma_1$
\State Calculate $r^j_{\mathrm{prox}} := 
    \frac{J_{\mathrm{nl}}(Z^{j}) - J_{\mathrm{nl}}(Z^{j+1})}{J_{\mathrm{nl}}(Z^{j}) - J_{\mathrm{lin}}(Z^{j+1}, Z^{j}, w_{\mathrm{prox}})}$
\If{$r^j_{\mathrm{prox}} \leq \beta_1$}
    \State Reject the solution $Z^{j+1}$
    \State $w_{\mathrm{prox}} \gets w_{\mathrm{prox}} \sigma_1$
\Else
    \State Accept the solution $Z^{j+1}$
    \If{$\beta_1 < r^j_{\mathrm{prox}} < \beta_2$}
        \State $w_{\mathrm{prox}} \gets w_{\mathrm{prox}} \sigma_2$
    \Else
        \State $w_{\mathrm{prox}} \gets w_{\mathrm{prox}} \sigma_3$
    \EndIf
\EndIf
\State \textbf{Return:} $w_{\mathrm{prox}}$
\end{algorithmic}
\end{algorithm}

The implementation of the prox-linear method can be further enhanced using the acceleration scheme described in \cite[Sec. 7]{drusvyatskiy2019efficiency}.
We use CVXPY \cite{diamond2016cvxpy} with either ECOS \cite{domahidi2013ecos} or MOSEK \cite{aps2019mosek} for modeling and solving \eqref{lpen_fd}. Customizable conic solvers such as the proportional-integral projected gradient (PIPG) algorithm \cite{yu2022proportional, yu2022extrapolated} could be used for efficient implementations.

\section{Numerical Results}  \label{sec:results}

\begin{figure}
\centerline{\includegraphics[scale=0.58]{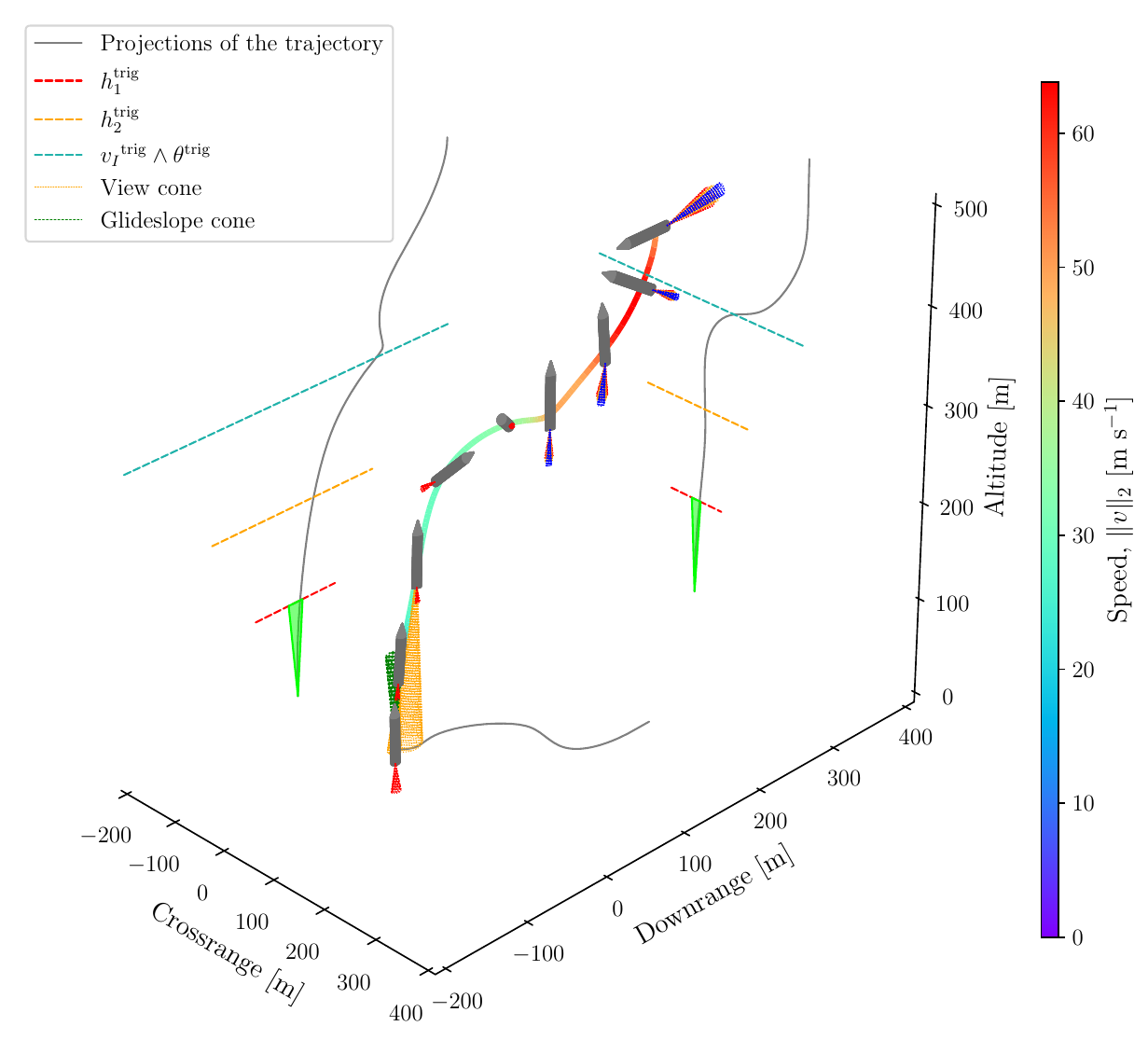}}
\caption{The landing trajectory of the rocket.}
\label{fig:rl_pos}
\end{figure}

\begin{figure}
\centerline{\includegraphics[scale=0.59]{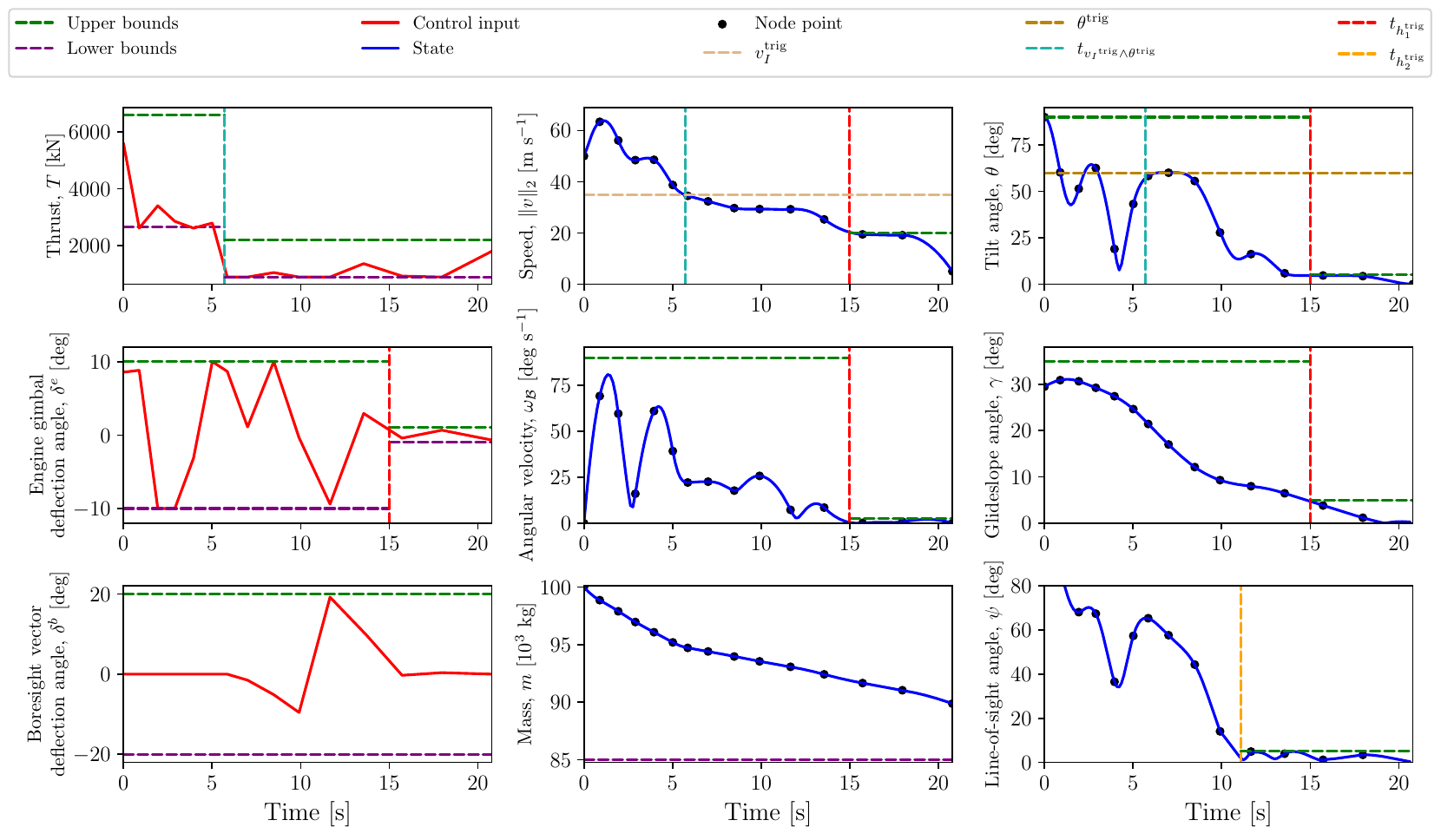}}
\caption{State and control inputs of the rocket during landing.}
\label{fig:rl_oth}
\end{figure}

The solution to the powered descent guidance problem with continuous-time compound state-triggered constraints is illustrated in Figures \ref{fig:rl_pos} and \ref{fig:rl_oth}. The system and simulation parameters are presented in Table \ref{tab:roc-lnd-look-param}. To improve the performance of the optimizer, affine scaling is applied to the state and control inputs. Initial guesses are set as follows: the state is interpolated linearly between initial and final values, the thrust force is set to $\frac{1}{2} \big( T^{\mathrm{stc}_1}_{\mathrm{max}} + T^{\mathrm{stc}_1}_{\mathrm{min}} \big)$ through the flight, gimbal deflection and azimuth angles are set to $0$, and the free-final-time is initialized at $21$ s.

At approximately $5.72$ s, the vehicle's speed and tilt angle drop below the specified thresholds, $v_{\mathcal{I}}^{\mathrm{trig}}$ and $\theta^{\mathrm{trig}}$, respectively with the thrust constraints satisfied immediately afterward. 
Around $11.07$ s, 
the altitude decreases below $h^{\mathrm{trig}}_2$, triggering the satisfaction of the line-of-sight constraint. 
Finally, the altitude decreases below $h^{\mathrm{trig}}_1$ around $15$ s, 
followed by the satisfaction of constraints on the gimbal deflection angle, speed, angular velocity, tilt angle, and the glideslope angle. 
All the conditions are met between consecutive node points and the corresponding constraints are satisfied in continuous-time immediately afterward.
The implementation is available at \url{https://github.com/UW-ACL/CT-cSTC}

\vspace{-0.25cm}
\begin{table}[!htpb]
\centering
\caption{}\label{tab:roc-lnd-look-param}
{\renewcommand{\arraystretch}{1.1}
\begin{minipage}[t]{0.45\linewidth}
\centering
{
\renewcommand{\arraystretch}{1.325}
\begin{tabular}{|c||c|}
\hline
Parameter & Value\\
\hline 
$K$ & $15$ \\
$g_0$ & $9.806$ m s$^{-2}$ \\
$\rho$ & $1.225$ kg m$^{-3}$ \\
$I_{\mathrm{sp}}$ & $330$ s \\
$J_{\mathcal{B}}(t)$ & $m(t) \mathrm{diag}([60, 60, 1.5])$ kg m$^2$ \\
$C_A$ & 
$\mathrm{diag}([0.4068, 0.4068, 0.0522])$ \\
$S_A$ & $545$ m$^{2}$ \\
$r_{\mathrm{cm},\mathcal{B}}$ & $(0, 0, -14)$ m \\
$r_{\mathrm{cp},\mathcal{B}}$ & $(0, 0, 3)$ m \\
$m_{\mathrm{i}}$ & $100000$ kg \\
$m_{\mathrm{dry}}$ & $85000$ kg \\
$r_{\mathrm{i}}$ & $(200, 200, 500)$ m \\
$r_{\mathrm{f}}$ & $(0, 0, 0)$ m \\
$v_{\mathrm{i}}$ & $(0, 0, -50)$ m s$^{-1}$ \\
$v_{\mathrm{f}}$ & $(0, 0, -5)$ m s$^{-1}$ \\
$q_{{\mathcal{B} \leftarrow \mathcal{I}}_\mathrm{i}}$ & $(\sqrt{2}, \sqrt{2}, 0, 0)$ \\
$q_{{\mathcal{B} \leftarrow \mathcal{I}}_\mathrm{f}}$ & $(1,0,0,0)$ \\
$\omega_{\mathcal{B}_\mathrm{i}}$ & $(0,0,0)$ deg $s^{-1}$ \\ 
$\omega_{\mathcal{B}_\mathrm{f}}$ & $(0,0,0)$ deg $s^{-1}$ \\
\hline
\end{tabular}
}
\end{minipage}%
\hfill
\begin{minipage}[t]{0.45\linewidth}
\centering
{\renewcommand{\arraystretch}{1.205}
\begin{tabular}{|c||c|}
\hline
Parameter & Value\\
\hline
$\omega_{\mathrm{max}}$ & $90$ deg $s^{-1}$ \\
$\theta_{\mathrm{max}}$ & $90$ deg \\ 
$\gamma_{\mathrm{max}}$ & $35$ deg \\
$\delta_{\mathrm{max}}^{\mathrm{e}}$ & $10$ deg \\
$\phi_{\mathrm{max}}^{\mathrm{e}}$ & $180$ deg \\
$\delta_{\mathrm{max}}^{\mathrm{b}}$ & $20$ deg \\
$\phi_{\mathrm{max}}^{\mathrm{b}}$ & $180$ deg \\
$h^{\mathrm{trig}}_1$ & $100$ m \\
$h^{\mathrm{trig}}_2$ & $200$ m \\
$v_{\mathcal{I}}^{\mathrm{trig}}$ & $35$ m s$^{-1}$ \\
$\theta^{\mathrm{trig}}$ & $60$ deg \\
$v_{\mathcal{I}}^{\mathrm{stc}}$ & $20$ m s$^{-1}$ \\
$\omega^{\mathrm{stc}}$ & $2.5$ deg s$^{-1}$ \\
$\theta^{\mathrm{stc}}$ & $5$ deg \\
$\gamma^{\mathrm{stc}}$ & $5$ deg \\
$\psi^{\mathrm{stc}}$ & $5$ deg \\
$\delta^{\mathrm{stc}}$ & $1$ deg s$^{-1}$ \\
$T^{\mathrm{stc}_1}_{\mathrm{max}}$ & $2$ $200$ $000$ kg m s$^{-2}$ \\
$T^{\mathrm{stc}_1}_{\mathrm{min}}$ & $880$ $000$ kg m s$^{-2}$ \\
$T^{\mathrm{stc}_2}_{\mathrm{max}}$ & $6$ $600$ $000$ kg m s$^{-2}$ \\
$T^{\mathrm{stc}_2}_{\mathrm{min}}$ & \hspace{0.83cm}  $2$ $640$ $000$ kg m s$^{-2}$ \hspace{0.83cm} \\ 
\hline
\end{tabular}
}
\end{minipage}
}
\end{table}
\vspace{-0.3cm}

\section{Conclusion and Future Works} \label{sec:conclusion}
This paper introduces a sequential convex programming (SCP) framework for solving the powered descent guidance problem with continuous-time satisfaction of state-triggered constraints. The framework ensures robust and efficient optimization by leveraging the smoothness, soundness, and completeness of temporal and logical specifications parameterized via the generalized mean-based smooth robustness measure (D-GMSR). The continuous-time successive convexification (CT-SCvx) method is applied, incorporating techniques such as path constraint reformulation, time-dilation, multiple shooting, exact penalty functions, and a convergence-guaranteed prox-linear SCP algorithm. The proposed approach is validated through numerical simulation, demonstrating its effectiveness and feasibility.

Future work could focus on developing a continuous-time formulation for temporal specifications such as \textit{eventually} and \textit{until}. Additionally, optimization algorithms tailored to leverage the structure of temporal and logical specifications may be designed to solve the resulting problems more efficiently.

\section*{Acknowledgments}
This research was supported in part by NASA Cooperative Agreement Grant 80NSSC24M0212,  AFOSR grant FA9550-20-1-0053, and ONR grant N00014-20-1-2288; Government sponsorship is acknowledged.

\bibliography{sample}

\end{document}